\documentclass[prd,preprintnumbers,floatfix,aps,nofootinbib,notitlepage,twocolumn,showpacs]{revtex4}

\usepackage{latexsym}
\usepackage{epsfig}
\usepackage{amssymb}

\newcommand{\lp}{\left(}
\newcommand{\rp}{\right)}
\newcommand{\lb}{\left[}
\newcommand{\rb}{\right]}
\newcommand{\LF}{\left(}
\newcommand{\RF}{\right)}
\newcommand{\LT}{\left[}
\newcommand{\RT}{\right]}
\newcommand{\p}{\partial}

\newcommand{\bea}{\begin{eqnarray}}
\newcommand{\eea}{\end{eqnarray}}
\newcommand{\non}{\nonumber\\}

\newcommand{\mx}{\mbox}
\newcommand{\mt}{\mathtt}
\newcommand{\mand}{\mx{ and }}

\newcommand{\stwo}{\sqrt{2}}
\newcommand{\sthree}{\sqrt{3}}

\newcommand{\Ra}{\Rightarrow}
\newcommand{\cA}{{\cal A}}

\newcommand{\cO}{{\cal O}}
\newcommand{\cP}{{\cal P}}

\newcommand{\fnl}{f_{\mt{NL}}}
\newcommand{\gnl}{g_{\mt{NL}}}
\newcommand{\ra}{\rightarrow}

\newcommand{\ba}{\begin{eqnarray}}
\newcommand{\ea}{\end{eqnarray}}
\newcommand{\be}{\begin{equation}}
\newcommand{\ee}{\end{equation}}

\newcommand{\al}{\alpha}
\newcommand{\bt}{\beta}
\newcommand{\ga}{\gamma}
\newcommand{\ka}{\kappa}

\newcommand{\la}{\lambda}
\newcommand{\za}{\zeta}

\newcommand{\Da}{\Delta}

\newcommand{\La}{\Lambda}

\begin{document}

\title{Phase transitions during Cyclic-Inflation and Non-gaussianity}

\author{Tirthabir Biswas}
\email{tbiswas@loyno.edu}
\affiliation{
Department of Physics,
Loyola University,
6363 St. Charles Avenue, Campus Box 92,
New Orleans, LA 70118, USA}
\author{Tomi Koivisto}
\email{tomik@astro.uio.no}
\affiliation{Institute of Theoretical Astrophysics, University of Oslo, N-0315 Oslo, Norway}
\author{Anupam Mazumdar}
\email{a.mazumdar@lancaster.ac.uk}
\affiliation{Consortium of particle physics and cosmology, Lancaster University, LA1 4YB, UK}

\date{\today}

\begin{abstract}
Typically cold inflation with positive vacuum energy density dilutes all matter except the quantum fluctuations which are stretched outside the Hubble patch during inflation. However the cyclic-inflation scenario, where the universe undergoes  many many asymmetric cycles leading to an overall exponential growth, operates in the presence of thermal matter. Thus, while some of the nice properties of cold inflation is  preserved, in cyclic inflationary models thermal excitations are continually produced and there exists a unique opportunity of imprinting interesting thermal history in CMBR. In particular, we will see that a phase transition can lead to observably large higher order nongaussianities as the critical temperature is approached, and that these are increasingly divergent with the order of the correlations.
\end{abstract}


\maketitle


\section{Introduction}
It is well known that our universe can inflate over many non-singular cycles in such a way that every cycle undergoes a slight expansion in the scale factor of the universe. If the universe undergoes many cycles which results in $50$-$60$ e-foldings of inflation, known as cyclic inflation (CI)~\cite{cyclic-inflation}, then it can explain the large scale homogeneity of the universe. For the operation of the cyclic inflation scenario one requires a negative potential energy, $-\La$, along with thermal matter, both relativistic  and non-relativistic, which are expected in any dynamical evolution.  For instance, the relativistic particles can come from the Standard Model  degrees of freedom and the cold dark matter sector (which would all be relativistic at the energy scale we are interested in). Of course, if the universe is locked in the negative potential energy phase, it will keep on inflating. However, as shown in~\cite{bkm-exit}, depending upon the nature of the potential, one can exit primordial inflation through a classical transition from negative to positive potential regions, see also~\cite{Lidsey:2006md,Nunes:2005ra,Lidsey:2004ef,felder} for the study and implementation of similar mechanisms. To our knowledge, CI model was the first  to realize a phenomenologically successful cosmology with negative potential energy~\footnote{For more recent efforts in this direction see~\cite{hartle}, though ghost and tachyonic instabilities may develop in this proposal \cite{Mithani:2013ed}. It is worth stressing that the nonsingular scenario of Refs. ~\cite{cyclic-inflation,bkm-exit} can be based on ghost-free gravity \cite{Biswas:2011ar}. See also \cite{Prokopec:2006yh} for a different attempt at nonlocal reconciliation of AdS and our universe.}, and may have interesting implications for String theory and supergravity which contains negative vacua~\cite{flux-compactification}.

In the  present paper we will argue that the origin of the ``relevant'' seed  perturbations  that were created during the cyclic phases, i.e. roughly speaking during the $6-7$ e-foldings of inflation that is seen in the cosmic microwave background (CMB) radiation,  are statistical thermal fluctuations in nature. In particular we will look at  the two point correlation function which is related to the power spectrum, $\cP_{\za}$, and the higher point correlation functions which are related to  the parameters $\fnl$ and $ \gnl$ characterizing the non-gaussian features in CMB. The best present constraints on the $\fnl$ come from the WMAP7 data and for the local type of nongaussianity imply $-10<\fnl<74$ \cite{Komatsu:2010fb}. For the $\gnl$, the constraints are considerably weaker and both CMB \cite{Vielva:2009jz} and large scale structure \cite{Desjacques:2009jb,Enqvist:2010bg} restrict it only to be less than about $10^5$ in magnitude.

In a companion paper~\cite{BBKM-thermal}, we provide a comprehensive discussion and derivation of how statistical fluctuations that exist in any thermal ensemble can source fluctuations in CMB, see also~\cite{param,Chen,Cai-thermal}. We will focus on how in the CI scenario the observations of the CMB could reveal to us novel  thermal features, such as phase transition, which is impossible to detect in the traditional cold inflationary scenarios. In particular, we will see that if the scale of the negative potential energy is close to the critical temperature of a phase transition, then we can hope to  find observable nongaussian signatures. Interestingly, the signal increases as one looks at higher order correlation functions. So for instance, while we find $\fnl$ to be too small to be detected by Planck, $\gnl$ as well as $\tau_{\mt{NL}}$ can be large enough to be observed. In addition we expect their amplitude to peak over a certain range of scales, which is a further very distinguishing prediction of the thermal phase transitions we explore here. 

So, let us now provide a brief summary of how the cyclic evolution works~\cite{cyclic-inflation,bkm-exit,cyclic-prediction}. Let us begin our story at the expanding phase of a given cycle with the universe containing some thermal matter along with a negative cosmological constant~\footnote{The generalization to a slowly evolving negative potential energy required to obtain the observed spectral tilt and the graceful exit is discussed in~\cite{cyclic-prediction} and~\cite{bkm-exit} respectively.}. As our universe expands, all the matter components dilute and is eventually canceled by the negative cosmological constant causing the universe to turnaround and start contracting. In Einstein's gravity, this leads to an eventual collapse, but, for instance, in an extension of General Relativity with infinite higher derivatives~\cite{BMS,BBMS,BKM,BKMV,alex} it is possible to resolve this cosmological singularity, and thus the contraction phase can nonsingularly transition into an expansion phase thereby completing the cycle.
An interesting fact about this class of infinite derivative theories is that in the ultraviolet regime it resolves the cosmological singularity \footnote{In general the infrared divergence of quantum fluctuations that could otherwise dominate over the thermal fluctuations may also be resolved in nonsingular cosmologies \cite{Koivisto:2010pj}.}, but in the far infrared the theory behaves exactly as that of the Einstein's theory of gravity. The bounce happens at a scale close to the  Planckian time $M_p^{-1}$, where $M_p$ is the reduced Planck mass~\footnote{For some of the other attempts to resolve the Big Bang/Big Crunch singularity through bounces when a high energy scale is reached, see~\cite{Ashtekar,Freese,Baum,Bojowald:2006da,Shtanov,Barragan:2009sq,Koivisto:2010jj}.}. Once the universe departs from the Planckian time scale, i.e. as the universe expands, we are far away from the bounce and the dynamics of the universe is  governed by Einstein's theory. Since, as will soon become apparent, all the interesting physics that we are going to focus on in this paper occurs near the turnaround which is far away from the bounce, the details of the bounce mechanism should not matter. Thus, while we do not pretend that our assumption about the smooth resolution of the Big Bang singularity  should be taken lightly, at least if such a resolution mechanism via bounce do exist in nature around the Planck scale, we fully expect our analysis to be independent of the new physics.

\begin{figure}
\includegraphics[width=7cm]{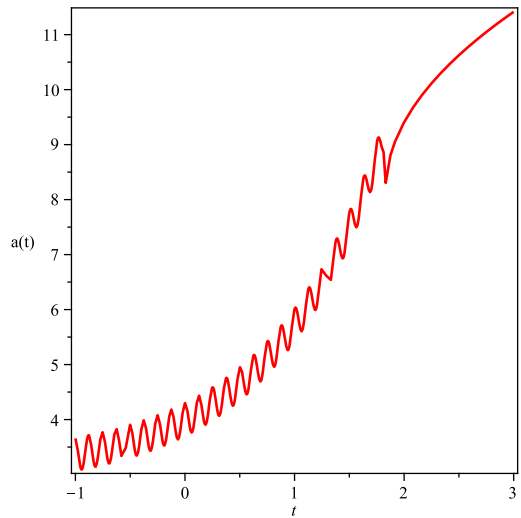}
\caption{\label{fig:ci}
Cyclic Inflation Scenario: The universe starts in a negative energy phase undergoing asymmetric cyclic evolution. There occurs a last cycle when the universe makes a transition from negative to positive energy phase followed by usual radiation dominated expansion.}
\end{figure}
Now, interactions between relativistic and non-relativistic species create entropy which can only increase monotonically thereby breaking the periodicity of the cyclic evolution. In fact, this was precisely what Tolman pointed out in the 1930's giving rise to Tolman's entropy problem~\cite{tolman1,tolman2}. However, this turns out to be a great asset for the CI scenario. As shown in~\cite{cyclic-inflation,cyclic-prediction}, entropy tends to increase by the same factor in every cycle, while the time period of the cycles remain a constant since it is governed by $|\La|$. This means that the universe must grow by the same factor, say by $(1+\ka)$, in every cycle giving rise to an overall inflationary growth, see Fig.~\ref{fig:ci}. Most of the advantages of standard slow-roll inflation is straight forwardly transferred to the CI scenario, including the production of nearly scale-invariant density fluctuations~\cite{cyclic-prediction}. Tolman's entropy problem also finds an elegant resolution if the universe is closed~\cite{emergent-cyclic}.  For more details of the cyclic inflationary scenario and how it can solve the different cosmological puzzles (flatness, horizon and homogeneity) the readers are referred to~\cite{cyclic-inflation,bkm-exit,cyclic-prediction}. In this paper our primary focus would be the computation of the various features in the primordial fluctuations that are present during the CI phase.

The paper is organized as follows. In section~\ref{tcis}, we review the evolution of perturbations during the cyclic inflation phase. The section~\ref{cmb} then focuses on the thermal fluctuations in this scenario, in particular we consider phase transitions and their impact on nongaussianity. Tensor modes are found to be insignificant in this context. Section \ref{conc} briefly concludes. The technical issue of the transition of the modes from sub- to super-horizon is treated in the appendix.

\section{Brief discussion on creating perturbations from cyclic Inflation}
\label{tcis}
In Refs.~\cite{cyclic-inflation,cyclic-prediction}, a new way of obtaining nearly scale-invariant fluctuations has been proposed in the context of cyclic inflationary models. The underlying mechanism is very similar to the standard inflationary scenarios, but the implementation is very different. To understand how the thermal fluctuations evolve it is useful to compare the cosmological time scale and the physical wavelength of fluctuations under consideration. Naively, if the physical wavelength is shorter than the cosmological scale (sub-Hubble phase), then the fluctuations are expected to be sourced by thermal statistical fluctuations in the fluid~\footnote{For a detailed discussion on the subject the reader is referred to our companion paper~\cite{BBKM-thermal}.}, but once the wavelengths become larger than the cosmological time scale, thermal correlations cannot be maintained and the coupled metric and fluid fluctuations are governed by the usual differential equations governing an ideal fluid. This transition is rather similar to the usual transition from the sub- to the super-Hubble phase in inflationary cosmology, after which the amplitude freezes.

\begin{figure}[htbp]
\begin{center}
\includegraphics[width=0.35\textwidth,angle=0,scale=0.8]{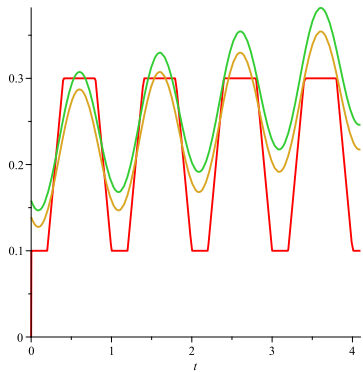}
\end{center}
\caption{The Last Exit: The curve in the red corresponds to the cosmological time-scale, while the green and the yellow curves represent the physical wavelengths of two different comoving modes. The two modes are initially in a Mixed phase but then they make their last exits after which they evolve as super-Hubble modes. While the green wavelength exits in the third cycle, the yellow curve has to wait for another cycle to make it's last exit. The two modes experiences identical background cosmology, the yellow curve only lagging behind by a cycle as compared to the green curve.
  \label{fig:lastexit}}
\end{figure}
In the traditional monotonically expanding or contracting universe the cosmological time scale is given by the Hubble radius. According to this reasoning all modes start out in a ``pure sub-Hubble'' phase where the physical wavelengths of fluctuations are always smaller than the cosmological scale, which has a lower bound given by the bounce scale:
\be
\tau_b\sim { M_p\over \sqrt{\rho_b}}\sim M_p^{-1}\ ,
\ee
where $\rho_b\sim M_p^4$ determines the total energy density at the time of bounce. This phase lasts until the physical wavelength of the given comoving mode at the bounce point remains smaller than $\tau_b$. With the gradual increase of the wavelengths though, due to the asymmetric growth in the cycles,  the modes start to undergo a much more complicated history. The modes now start out sub-Hubble  at the turnaround point but exits the Hubble radius in the contracting phase. It re-enters the sub-Hubble phase near the bounce only to exit the Hubble radius shortly after the bounce. Finally, they  cross back into the sub-Hubble domain in the expanding branch.

This cyclic pattern of evolution continues for a while,  but again as the universe undergoes inflationary growth over many cycles, the physical wavelength keeps increasing, so that the times that a given mode spends in the sub-Hubble phases around the turnarounds and the bounces become shorter and shorter. Eventually there comes a cycle when the sub-Hubble phases are shorter than the physical wavelength of the modes. From  this cycle on, although technically the modes do undergo sub-Hubble evolutions, they are ineffective in establishing thermal correlations, and the modes effectively remains super-Hubble till the rest of the duration of the CI phase. We call this the ``last exit''. Thus the CI scenario mimics the standard inflationary mechanism of freezing the amplitude of perturbations after the Hubble crossing.

Approximately it is clear that the causal time scale in our scenario is given by the time period of the cycle
\be
\tau=\bt{ M_p\over \sqrt{\La}}\ ,
\label{taub}
\ee
where $\bt\sim \cO(1)$ constant, and therefore the modes must be exiting when their wavelengths $\sim \tau$. In other words, the modes exit very near the turnaround energy density, in the appendix, we provide a quantitative estimate of this time scale. A more physical way of thinking about the whole process is simply to realize that the cosmological time scale does not correspond to the Hubble radius near turnarounds and bounces, but is rather controlled by $\tau$ and $\tau_b$ respectively, which in turn are determined by the typical energy densities near these regions. The fact that the Hubble scale does not always correspond to the cosmological scale has indeed been discussed before~\cite{kinney}.

We now come to a very important point: Suppose $k_1$ is a mode which is the first to make the last exit in a given cycle, then once the wave number increases by a factor $(1+\ka)$, the fluctuations can no longer undergo the last exit in that given cycle because they re-enter the thermal sub-Hubble evolution during the following turnaround, see Fig.~\ref{fig:lastexit}. They therefore,  must wait for the next cycle for their last exit. What this means is that, if $\ka$ is a small parameter then the last exits always occur at approximately the same energy densities, $\sim \La$, near the turnarounds. We know from our study of perturbations that the energy density(or equivalently the temperature in our case) at the exit controls the amplitude of fluctuations that we eventually see in the CMB. This therefore explains why cyclic inflation model gives rise to an approximate scale-invariant spectrum, preserving one of the most robust predictions of inflation theory.

From the above discussion, a rather distinguishing feature of the cyclic inflation model also emerges which was  highlighted in~\cite{cyclic-prediction}. The fact that in a given cycle only a certain range of modes make their last exit means that there should exist small wiggles in the primordial spectrum of the form
\be\label{final}
\cP_{\Phi}=\cA^2S(k)
\ee
where $\cA$ is strictly a constant if we have a negative cosmological constant but will vary slowly if we have a $V(\phi)$ giving rise to an overall tilt in the spectrum, while $S(k)$  is a periodic function in $\ln k$:
\be
S(k(1+\ka))=S(k)\,.
\ee
The precise shape of $S(k)$ is determined by how the temperature rises as the modes exit in a give cycle, which in turn depends on the thermodynamic properties of the fluid. While our previous analysis provided some tantalizing hints for such oscillatory features, the Planck data would be able to shed more light. For us, the important point is that the oscillatory feature can be used in conjunction with the nongaussian features, that we will soon discuss,  to differentiate our model from other inflationary scenarios that also produce large nongaussianities.
\section{Statistical thermal fluctuations \& CMB}
\label{cmb}
In many extensions of the Standard model, we expect our universe to have undergone phase transitions at high temperatures, these include Grand Unified theories (supersymmetric and non-supersymmetric) as well as String theory. Could these phase transitions have left any observable imprint in our sky? In this section we will provide a simple example of thermal phase transition which can leave detectable signals in the CMB. First the phase transition parameterisation will be described, then we will apply the results
of the companion paper~\cite{BBKM-thermal} to compute the resulting bi- and trispectrum. Finally gravitational wave production is discussed briefly.

\subsection{Phase Transition}

Phase transitions are characterized by critical exponents. For our purpose the most relevant is the one associated with the heat capacity, commonly denoted by $\al$:
\be
\lim_{T\ra T_c}C_V\sim\left|{T-T_c\over T_c}\right|^{\al}\,,
\label{critical}
\ee
$\al=-1$ corresponds to a first order phase transition, where the heat capacity diverges. We are going to restrict ourselves to higher order transitions with $\al>-1$, and consider the case when the temperature of the last exit is below the critical temperature. A prototype partition function corresponding to critical behavior such as (\ref{critical}) is given by
\bea
p &= &{T\ln Z\over V} =  \frac{1}{3}gT^4 - \Lambda \non
&+& \mu T_0^4 \lb \lp 1-\frac{T}{T_c}\rp^{\al+2}-1\rb\,.
\eea
The  two terms in the first line correspond to the pressure coming from the usual massless degrees of freedom, and the negative cosmological constant that is required in the CI model to provide the turnarounds.
The term in the second parentheses contain the physics of the phase transition; incorporating some multiplicative modulating functions of temperature should not change the results in any significant way as we are only going to be interested in temperatures very close to the critical temperature, $T_c$. The parameter $g$ is the number of effective massless degrees of freedom times $(\pi^2/30)$, and the parameter $\mu$ controls the contribution of the thermal density component undergoing the phase transition; we consider the regime $g \gg \mu$.
The temperature at the turnover is $T_0$, and it provides the overall energy scale. In terms of these parameters, the cosmological constant is given by
\bea
\Lambda &  = & \rho(T\ra 0) = -p(T \ra 0)  =  -T_0^4\Big[ g + \mu \non
& - & \mu\lp 1+ \frac{T_0}{T_c}\rp^{\al+1}\lp 1+(1+\al)\frac{T_0}{T_c}\rp\Big]\,,
\eea
and as mentioned it is negative in the CI scenario.

\begin{figure}
\includegraphics[width=8.9cm]{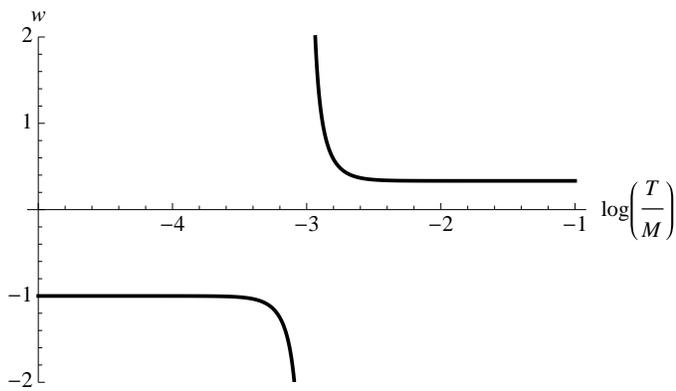}
\caption{\label{fig:EoS}
The equation of state of the fluid as a function of temperature when $T_c=0.1M_p$ and $T_0=0.01M_p$. We use $\al=0.1$, $g=100$ and $\mu=0.1$. At high temperatures, the radiation dominates and $w=1/3$, at low temperatures the cosmological constant is important and $w=-1$. The divergence occurs at $T=T_0$ where the density vanishes.}
\end{figure}

Noting the thermodynamic relation
\be
\rho(T)=-{1\over V}{\p (\ln Z)\over \p \bt}=T{d p(T)\over d T}-p(T)\,,
\ee
one obtains
\bea
\rho & = & gT^4+\Lambda \non
& - & \mu T_0^4\lb \lp 1-\frac{T}{T_c}\rp^{\al+1}\lp 1+ (1+\al)\frac{T}{T_c}\rp-1\rb\,, \label{density}\ , \\
\frac{C_V}{V} & = & 4g T^3+\mu\frac{T_0^4 T}{T_c^2}(1+\al)(2+\al)\lp 1-\frac{T}{T_c}\rp^{\al}\,.
\eea
One readily checks that this is consistent with the density vanishing at the turnaround $T=T_0$ and
reducing to the cosmological term at the zero temperature. We plot the equation of state in Fig.~\ref{fig:EoS}.

\subsection{Non-Gaussianity}
The expressions for calculating the power spectrum and the nongaussianity parameters from statistical fluctuations
are derived in our companion paper~\cite{BBKM-thermal}.
Here we summarize the results:
\bea
\cP_{\zeta} & = & \gamma^2\sqrt{3}A^2(T)\frac{T^2\rho'}{M_p^3\sqrt{\rho}}\,, \label{spectrum} \\
f_{NL} & = &
 \lp\frac{1}{\gamma}\rp \frac{1}{A(T)}\lb \frac{5\rho\lp 2\rho' + T\rho''\rp}{24T\lp \rho'\rp^2}\RT\,, \label{f_nl} \\
g_{NL} & = &
 \lp \frac{1}{\ga^2} \rp \frac{1}{A^2(T)}
\lb \frac{25\rho^2\lb 3\lp \rho' + T\rho''\rp + T^2\rho'''\RT}{243T^2\lp\rho'\rp^3} \RT \label{g_nl}\,, \non
\eea
where
\bea
A(T) &\equiv& \frac{9 + 3w + 2r}{6(1+w)} \label{A}\\
r&\equiv& -{3\over 2}\LT1+\frac{\lp1+w\rp\rho\lp2\rho'+T\rho''\rp}{T{\rho'}^2}\,\RT \label{r}\\
w&=&{p\over \rho}\mand \gamma=2\stwo\pi^{3/4}\approx 6.7\ .
\eea
We have used the value of $\ga$ that corresponds to a Gaussian window function~\cite{BBKM-thermal}, and have set $\Omega=1$ since  (\ref{density}) already incorporated the energy coming from the cosmological constant, radiation and the thermal matter undergoing phase transition~\footnote{In the CI scenario, one also requires the presence of non-relativistic species~\cite{cyclic-prediction}, but it's always present in small amounts, and in fact, decays before the onset of the contraction phase.}.

Before trying to compute and plot these different quantities, let us try to understand the physics with the help of some approximations. To begin with we have 6 parameters: $T_e,T_c,T_0,\mu,g,\al$, where $T_e$ is the exit temperature. However, not all of these parameters are independent or important. First, we can choose $g$ to be the number appropriate for Standard Model, $g\sim (\pi^2/30)\cO(100)$. The value we use for numerics is $g=100$. We note in passing that one of the virtues of the CI model is that there is no need for reheating and the universe could have only  contained the Standard Model degrees of freedom all along.
Also, our computation of the exit temperature in terms of $T_0$ in the appendix was done for radiative matter sources and since $\mu\ll g$, the result remains applicable to our case; as also seen in Fig.~\ref{fig:EoS}, the radiative matter dominates the system when $T>T_0$. Hence it is justified to fix
\be \label{Trel}
T_0 = 0.8 T_e\,.
\ee
To consider a definite and relevant phase transition, we can use the critical exponent corresponding to $\phi^4$ theory. It has been computed/measured using various methods, and the results converge to $\al \approx 0.1$, see for instance~\cite{alpha}. 
This is the value we use for numerical examples unless otherwise specified.

\begin{figure}
\includegraphics[width=8.9cm]{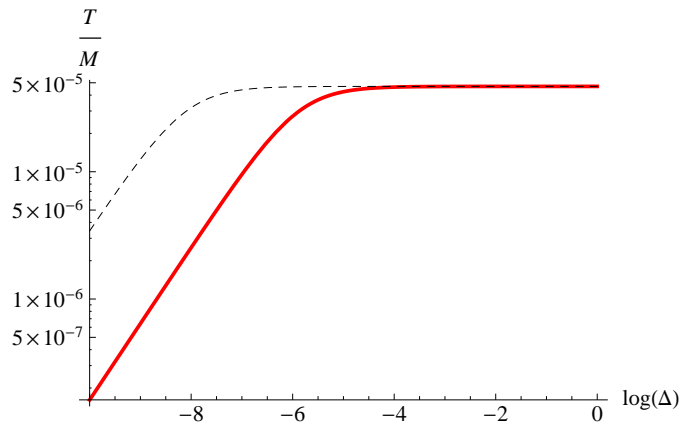}
\caption{\label{TperM}
The turnover temperature $T_0$ of the fluid as a function of  $\Da$.  For the red solid line, $\mu=0.1$ and for black dashed line, $\mu=1/1000$.}
\end{figure}

We are thus left with three parameters: $ T_0,\mu$ and $T_c$. Instead of the latter we shall refer to
\be
\Da\equiv (T_e-T_c)/T_c\,.
\ee
However, the amplitude of CMB fluctuations are known and this can be used to constrain one of the parameters, a convenient choice is $T_0$, which gives the overall scale of the cycles. We show in Fig.~\ref{TperM} solutions for the temperature solved by matching the amplitude of the spectrum with the observed WMAP value. To conclude, our predictions depend upon two parameters: the contribution of the matter subject to phase transition $\mu$, and the proximity to the critical temperature, $\Da$.

\begin{figure}
\includegraphics[width=8.9cm]{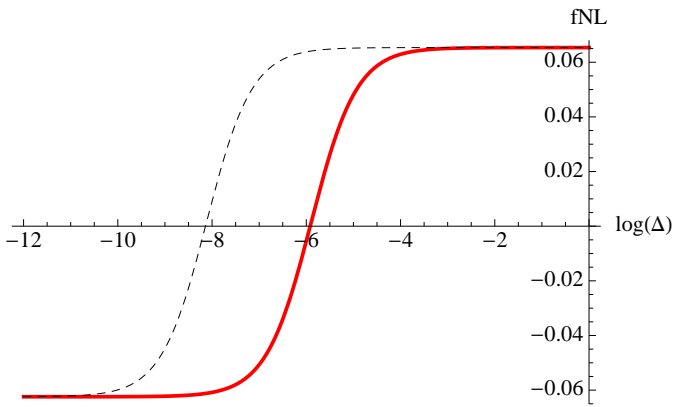}
\caption{\label{fNL}
Nongaussianity $f_{NL}$ as a function of  $\Da$.  For the red solid line, $\mu=0.1$ and for black dashed line, $\mu=1/1000$.}
\end{figure}

The essential feature of our results is large higher order nongaussianity.
They contain more and more derivatives of $\rho(T)$ and therefore become more and more divergent as $T$ approaches $T_c$.
Accordingly, the $f_{NL}$ parameter remains typically small, as seen in Fig.~\ref{fNL}. The parameter approaches a constant as $T\ra T_c$ and another constant far away from the critical temperature.
The next order nongaussianity $g_{NL}$ is already much more interesting: the parameter diverges as $T\ra T_c$.  To understand this, note that the leading term in this limit, from (\ref{g_nl}), is due to the third derivative of the density that goes like  $\rho'''\sim \Da^{\al-2}$.
However, one should also take into account the divergent terms in the denominator,  which is given by $A^2\sim r^2\sim \rho''^2\sim \Da^{2(1-\al)}$, as seen from Eqs.(\ref{A},\ref{r}). The result is that the nongaussianity diverges as $g_{NL}\sim \Da^{-\al}$. 

\begin{figure}
\includegraphics[width=8.9cm]{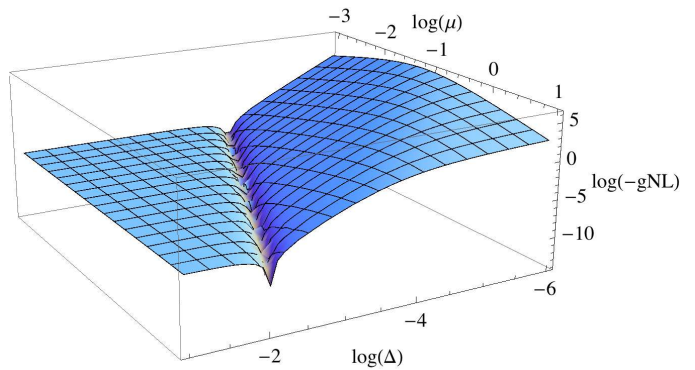}
\caption{\label{gNL}
Nongaussianity $g_{NL}$ as a function of  $\Da$ and $\mu$. The parameter is negative and diverges as $T\ra T_c$.}
\end{figure}

We illustrate this in the Fig.~\ref{gNL} where the nongaussianity is logarithmically plotted as a function of $\mu$ and $\Da$. While the fact that we only get an interesting signal for $\Delta \lesssim 10^{-4}$ may appear to be a fine-tuning, that is not so. We remind the readers that to be consistent with the observed spectral tilt of the spectrum $T_{0}$ and hence $T_{e}$ has to vary. For $n_s\sim 0.95$, $T_e$ varies by approximately $30\%$, which means as long as $T_c$ lies within this range, we can observe a signal. In fact based on our results, it seems that rather than having a broad $\gnl$ spectrum, we expect a sharp peak in $k$, which is another distinctive feature of the thermal signal. The higher order nongaussianities will peak at the wavelengths that cross the horizon as the phase transition occurs.

\begin{figure}
\includegraphics[width=8.9cm]{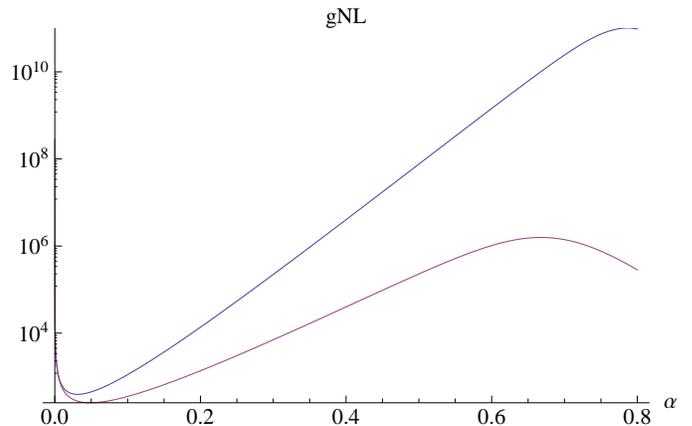}
\caption{\label{gNLa}
Nongaussianity $g_{NL}$ as a function of  the critical exponent $\al$ when $\mu=1$. For the upper line, $\Delta=10^{-15}$ and for the lower one, $\Delta=10^{-10}$. The monotonic scaling behavior comes apparent only very near the transition.}
\end{figure}

The dependence on hte critical exponent very near the transition is depicted in Fig.~\ref{gNLa}.
In brief, for positive $\al$, whenever we are close enough to the critical temperature, the $g_{NL}$ is large. In general, the nongaussianity corresponding to the $N$-point function diverges near the critical temperature as $(1-T/T_c)^{\al(3-N)}$, so the higher order nongaussianities can be significantly larger than the lower order ones. This property stems from the diverging behavior of the higher derivatives of the density $\rho$ and  can provide a smoking-gun signal for these models.

\subsection{Gravity Waves}

\begin{figure}
\includegraphics[width=8.9cm]{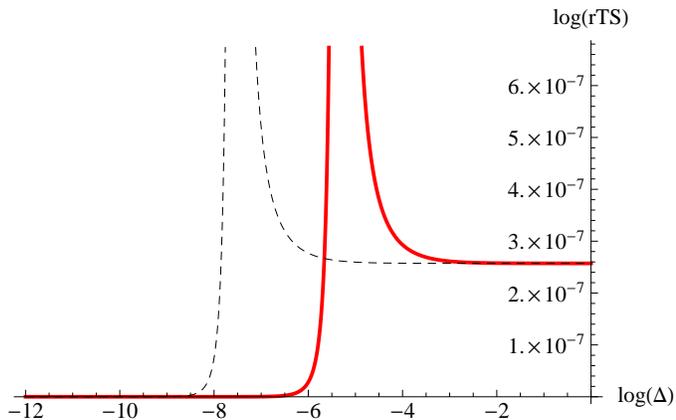}
\caption{\label{rTS}
The scalar-to-tensor ratio as a function of  $\Da$.  For the red solid line, $\mu=0.1$ and for black dashed line, $\mu=1/1000$.}
\end{figure}

To close the section of CMB constraints, let us investigate the gravitational wave production in the set-up. For extensive thermal systems, there are no off-diagonal stress energy components and as a result, there are no source terms for the gravitational waves. The gravity wave spectrum is therefore governed by the quantum vacuum fluctuations similar to what happens  in the standard scalar field driven inflationary paradigm assuming that the gravity is indeed quantum~\cite{amjad}. The gravity wave power spectrum is thus given by
\be
\cP_h={1\over 4\pi^3}\LF{H\over M_p}\RF^2 = {1\over 12\pi^3}{\rho\over M_p^4}\ ,
\ee
where $\rho$ corresponds to the energy density of the thermal fluid at the time of the last exit. Therefore the tensor to scalar ratio is given by
\be
r_{t/s}\equiv {\cP_{h}\over \cP_{\za}}={1\over 3\sqrt{3}\pi^3\ga^2}\LF{1+w\over 2+w+r}\RF^2{\rho^{3\over 2} \over T^2 M_p}{\over }\LF{ \p \rho\over   \p T}\RF^{-1}\ .
\ee
Again the expression has to be evaluated at the Hubble crossing, ${k=aH}$. The results are plotted in Fig.~\ref{rTS}. We see that the amplitude of the tensor modes is typically too small to be observed. For particular parameter combinations though, the scalar-to-tensor ratio diverges. However, the cyclic inflation to occur sufficiently near these singular points to obtain appreciable amount of gravitational waves would seem to require considerable fine-tuning. Hence we regard negligible gravitational wave contribution a robust prediction of the scenario.



\section{Conclusions}
\label{conc}

We computed the different features of CMB in the cyclic inflation scenario by applying the formalism developed for cosmological perturbations arising from statistical thermal fluctuations  in our companion paper~\cite{BBKM-thermal}.

In particular we explored the implications of phase transitions during the inflationary expansion.
We found that the nongaussianity corresponding to the $N$-point function of thermal fluctuations diverge near the critical temperature $T_c$ as $(1-T/T_c)^{\al(3-N)}$, where $\al$ is the critical exponent. The higher order correlations will be drastically amplified over a given range of scales. This generic feature is an imprint quite unique to the physics of thermal phase transitions. Observationally the higher order correlations are much more difficult to constrain, and since they are not typically huge in cold inflation models, the attention has been almost exclusively upon the bi- and trispectrum. However, as we predict that the amplitude of the signal is increasingly divergent for higher order nongaussianities, our findings can motivate the study of previously neglected completely new signatures.

Coupled with the firm prediction of negligible tensor amplitude and subtle wiggly features in the spectra~\cite{cyclic-prediction}, the structure of nongaussianities can provide a robust test of the cyclic inflation scenario.

\acknowledgements{TK  is supported by the Research Council of Norway. TB's research has been supported by the LEQSF(2011-13)-RD-A21 grant from the Louisiana Board of Regents. AM is supported by STFC-grant ST/J000418/1.}

\appendix
\section{Temperature of the Last Exit}
While intuitively it is clear that once the wavelengths become larger than the period of the oscillations, the modes should essentially behave as zero modes and follow the super-Hubble evolution, in any nonsingular cyclic model there are always two points within a cycle, the bounce and the turnaround, where the Hubble radius is infinite. So, according to the naive definition, all modes are back in the sub-Hubble phase! Of course, the crucial point is that larger the wavelength of the modes, the shorter the time that they spent in this sub-Hubble phase. Very conservatively, these modes need to spend at least a minimum of $\Da t\sim \la/c$  time  to have any ``sub-Hubble'' evolution at all. For instance, if the time the mode spends is less than $\la/c$, there is no way any thermal correlation could be established stretching a wavelength $\la$. Thus, although technically, all modes are sub-Hubble during the turnaround and the bounce, that is of little relevance. Once a mode reaches a particular physical wavelength they make their ``last exit'' and start behaving as superHubble modes for the cycles to follow. Let us in fact calculate this wavelength/temperature of the last exit point.

Let us denote the scale factor and temperature when a given mode, $k$, exits the Hubble radius in the contraction phase as $a_k$ and $T_k$ respectively:
\be
a_kH(a_k)=k
\label{Hubble-crossing}
\ee
Next, let us calculate the time that this mode spends in the sub-Hubble phase. Ignoring any asymmetry, this is given by
\be
\Da t =\int dt =2\int_1^{a_{k}} {da\over a H}
\ee
where we have chosen the convention that $a=1$ corresponds to the turnaround in question, so that $H(1)=0$. The smaller the $k$, the closer $a_k$ is to one, and therefore shorter is $\Da t$, as expected. Now, in order for any thermal correlation to re-establish $\Da t>\la=a/k$. Clearly then there exists a value of $k$ below which this condition can never be satisfied. The largest $k$ for which thermal correlation can be reestablished is then given by
\be
\Da t = {a_k\over k}
\label{exit-condition}
\ee
The above is an implicit equation for $k$ via (\ref{Hubble-crossing}). Using the above equation, for any given thermodynamic fluid, one can calculate the exit temperature, $k$, and $a_k$. Let us now demonstrate this for radiation.

For radiation in presence of a negative cosmological constant, we have
\bea
H&=&{\sqrt{\La}\over \sthree M_p}\sqrt{a^{-4}-1}\non
\Ra k&=&a_k{\sqrt{\La}\over \sthree M_p}\sqrt{a_k^{-4}-1}={\sqrt{\La}\over \sthree M_p}\sqrt{a_k^{-2}-a_k^2}\non
\mand \Da t  &= &{\pi\over 2}-\tan^{-1}\LF{a_k^2\over \sqrt{1-a_k^4}}\RF
\eea
(\ref{exit-condition}) can now be solved numerically to obtain $a_e$, the scale factor at which the last exit starts to occur. Approximately we find
\be
a_e\approx 0.8\Ra T_e\approx 1.25 T_0
\ee
where $T_0$ is the temperature at the turnaround point. More physically, all those modes whose physical wavelength at the turnaround point is larger than a certain wavelength,
\be
\la> {\sthree M_p\over \sqrt{\La}}{a_e\over \sqrt{1-a_e^4}}\approx 1.04\tau\ ,
\ee
they never undergo any thermal sub-Hubble evolution at all, they must have had their last exits in some previous cycle.

One can carry out a similar analysis during the bounce, since the time scale of the bounce is much shorter than the turnaround, any mode which doesn't undergo any sub-Hubble phase near the turnaround also has no sub-Hubble evolution during the bounce.

\bibliography{cyclicrefs}

\begin{thebibliography}{39}
\expandafter\ifx\csname natexlab\endcsname\relax\def\natexlab#1{#1}\fi
\expandafter\ifx\csname bibnamefont\endcsname\relax
  \def\bibnamefont#1{#1}\fi
\expandafter\ifx\csname bibfnamefont\endcsname\relax
  \def\bibfnamefont#1{#1}\fi
\expandafter\ifx\csname citenamefont\endcsname\relax
  \def\citenamefont#1{#1}\fi
\expandafter\ifx\csname url\endcsname\relax
  \def\url#1{\texttt{#1}}\fi
\expandafter\ifx\csname urlprefix\endcsname\relax\def\urlprefix{URL }\fi
\providecommand{\bibinfo}[2]{#2}
\providecommand{\eprint}[2][]{\url{#2}}

\bibitem[{\citenamefont{Biswas and Mazumdar}(2009)}]{cyclic-inflation}
\bibinfo{author}{\bibfnamefont{T.}~\bibnamefont{Biswas}} \bibnamefont{and}
  \bibinfo{author}{\bibfnamefont{A.}~\bibnamefont{Mazumdar}},
  \bibinfo{journal}{Phys.Rev.} \textbf{\bibinfo{volume}{D80}},
  \bibinfo{pages}{023519} (\bibinfo{year}{2009}), \eprint{0901.4930}.

\bibitem[{\citenamefont{Biswas et~al.}(2011)\citenamefont{Biswas, Koivisto, and
  Mazumdar}}]{bkm-exit}
\bibinfo{author}{\bibfnamefont{T.}~\bibnamefont{Biswas}},
  \bibinfo{author}{\bibfnamefont{T.}~\bibnamefont{Koivisto}}, \bibnamefont{and}
  \bibinfo{author}{\bibfnamefont{A.}~\bibnamefont{Mazumdar}}
  (\bibinfo{year}{2011}), \eprint{1105.2636}.

\bibitem[{\citenamefont{Lidsey and Mulryne}(2006)}]{Lidsey:2006md}
\bibinfo{author}{\bibfnamefont{J.~E.} \bibnamefont{Lidsey}} \bibnamefont{and}
  \bibinfo{author}{\bibfnamefont{D.~J.} \bibnamefont{Mulryne}},
  \bibinfo{journal}{Phys.Rev.} \textbf{\bibinfo{volume}{D73}},
  \bibinfo{pages}{083508} (\bibinfo{year}{2006}), \eprint{hep-th/0601203}.

\bibitem[{\citenamefont{Nunes}(2005)}]{Nunes:2005ra}
\bibinfo{author}{\bibfnamefont{N.~J.} \bibnamefont{Nunes}},
  \bibinfo{journal}{Phys.Rev.} \textbf{\bibinfo{volume}{D72}},
  \bibinfo{pages}{103510} (\bibinfo{year}{2005}), \eprint{astro-ph/0507683}.

\bibitem[{\citenamefont{Lidsey et~al.}(2004)\citenamefont{Lidsey, Mulryne,
  Nunes, and Tavakol}}]{Lidsey:2004ef}
\bibinfo{author}{\bibfnamefont{J.~E.} \bibnamefont{Lidsey}},
  \bibinfo{author}{\bibfnamefont{D.~J.} \bibnamefont{Mulryne}},
  \bibinfo{author}{\bibfnamefont{N.}~\bibnamefont{Nunes}}, \bibnamefont{and}
  \bibinfo{author}{\bibfnamefont{R.}~\bibnamefont{Tavakol}},
  \bibinfo{journal}{Phys.Rev.} \textbf{\bibinfo{volume}{D70}},
  \bibinfo{pages}{063521} (\bibinfo{year}{2004}), \eprint{gr-qc/0406042}.

\bibitem[{\citenamefont{Felder et~al.}(2002)\citenamefont{Felder, Frolov,
  Kofman, and Linde}}]{felder}
\bibinfo{author}{\bibfnamefont{G.~N.} \bibnamefont{Felder}},
  \bibinfo{author}{\bibfnamefont{A.~V.} \bibnamefont{Frolov}},
  \bibinfo{author}{\bibfnamefont{L.}~\bibnamefont{Kofman}}, \bibnamefont{and}
  \bibinfo{author}{\bibfnamefont{A.~D.} \bibnamefont{Linde}},
  \bibinfo{journal}{Phys.Rev.} \textbf{\bibinfo{volume}{D66}},
  \bibinfo{pages}{023507} (\bibinfo{year}{2002}), \eprint{hep-th/0202017}.

\bibitem[{\citenamefont{Hartle et~al.}(2012)\citenamefont{Hartle, Hawking, and
  Hertog}}]{hartle}
\bibinfo{author}{\bibfnamefont{J.~B.} \bibnamefont{Hartle}},
  \bibinfo{author}{\bibfnamefont{S.}~\bibnamefont{Hawking}}, \bibnamefont{and}
  \bibinfo{author}{\bibfnamefont{T.}~\bibnamefont{Hertog}}
  (\bibinfo{year}{2012}), \eprint{1205.3807}.

\bibitem[{\citenamefont{Mithani and Vilenkin}(2013)}]{Mithani:2013ed}
\bibinfo{author}{\bibfnamefont{A.~T.} \bibnamefont{Mithani}} \bibnamefont{and}
  \bibinfo{author}{\bibfnamefont{A.}~\bibnamefont{Vilenkin}}
  (\bibinfo{year}{2013}), \eprint{1302.0568}.

\bibitem[{\citenamefont{Biswas et~al.}(2012{\natexlab{a}})\citenamefont{Biswas,
  Gerwick, Koivisto, and Mazumdar}}]{Biswas:2011ar}
\bibinfo{author}{\bibfnamefont{T.}~\bibnamefont{Biswas}},
  \bibinfo{author}{\bibfnamefont{E.}~\bibnamefont{Gerwick}},
  \bibinfo{author}{\bibfnamefont{T.}~\bibnamefont{Koivisto}}, \bibnamefont{and}
  \bibinfo{author}{\bibfnamefont{A.}~\bibnamefont{Mazumdar}},
  \bibinfo{journal}{Phys.Rev.Lett.} \textbf{\bibinfo{volume}{108}},
  \bibinfo{pages}{031101} (\bibinfo{year}{2012}{\natexlab{a}}),
  \eprint{1110.5249}.

\bibitem[{\citenamefont{Prokopec}(2006)}]{Prokopec:2006yh}
\bibinfo{author}{\bibfnamefont{T.}~\bibnamefont{Prokopec}}
  (\bibinfo{year}{2006}), \eprint{gr-qc/0603088}.

\bibitem[{\citenamefont{Douglas and Kachru}(2007)}]{flux-compactification}
\bibinfo{author}{\bibfnamefont{M.~R.} \bibnamefont{Douglas}} \bibnamefont{and}
  \bibinfo{author}{\bibfnamefont{S.}~\bibnamefont{Kachru}},
  \bibinfo{journal}{Rev. Mod. Phys.} \textbf{\bibinfo{volume}{79}},
  \bibinfo{pages}{733} (\bibinfo{year}{2007}), \eprint{hep-th/0610102}.

\bibitem[{\citenamefont{Komatsu et~al.}(2011)}]{Komatsu:2010fb}
\bibinfo{author}{\bibfnamefont{E.}~\bibnamefont{Komatsu}} \bibnamefont{et~al.}
  (\bibinfo{collaboration}{WMAP Collaboration}),
  \bibinfo{journal}{Astrophys.J.Suppl.} \textbf{\bibinfo{volume}{192}},
  \bibinfo{pages}{18} (\bibinfo{year}{2011}), \eprint{1001.4538}.

\bibitem[{\citenamefont{Vielva and Sanz}(2010)}]{Vielva:2009jz}
\bibinfo{author}{\bibfnamefont{P.}~\bibnamefont{Vielva}} \bibnamefont{and}
  \bibinfo{author}{\bibfnamefont{J.}~\bibnamefont{Sanz}},
  \bibinfo{journal}{Mon.Not.Roy.Astron.Soc.} \textbf{\bibinfo{volume}{404}},
  \bibinfo{pages}{895} (\bibinfo{year}{2010}), \eprint{0910.3196}.

\bibitem[{\citenamefont{Desjacques and Seljak}(2010)}]{Desjacques:2009jb}
\bibinfo{author}{\bibfnamefont{V.}~\bibnamefont{Desjacques}} \bibnamefont{and}
  \bibinfo{author}{\bibfnamefont{U.}~\bibnamefont{Seljak}},
  \bibinfo{journal}{Phys.Rev.} \textbf{\bibinfo{volume}{D81}},
  \bibinfo{pages}{023006} (\bibinfo{year}{2010}), \eprint{0907.2257}.

\bibitem[{\citenamefont{Enqvist et~al.}(2011)\citenamefont{Enqvist, Hotchkiss,
  and Taanila}}]{Enqvist:2010bg}
\bibinfo{author}{\bibfnamefont{K.}~\bibnamefont{Enqvist}},
  \bibinfo{author}{\bibfnamefont{S.}~\bibnamefont{Hotchkiss}},
  \bibnamefont{and} \bibinfo{author}{\bibfnamefont{O.}~\bibnamefont{Taanila}},
  \bibinfo{journal}{JCAP} \textbf{\bibinfo{volume}{1104}}, \bibinfo{pages}{017}
  (\bibinfo{year}{2011}), \eprint{1012.2732}.

\bibitem[{\citenamefont{Brandenberger et~al.}(2013)\citenamefont{Brandenberger,
  Biswas, Koivisto, and Mazumdar}}]{BBKM-thermal}
\bibinfo{author}{\bibfnamefont{R.}~\bibnamefont{Brandenberger}},
  \bibinfo{author}{\bibfnamefont{T.}~\bibnamefont{Biswas}},
  \bibinfo{author}{\bibfnamefont{T.}~\bibnamefont{Koivisto}}, \bibnamefont{and}
  \bibinfo{author}{\bibfnamefont{A.}~\bibnamefont{Mazumdar}}
  (\bibinfo{year}{2013}), \eprint{1302.xxxx}.

\bibitem[{\citenamefont{Magueijo and Singh}(2007)}]{param}
\bibinfo{author}{\bibfnamefont{J.}~\bibnamefont{Magueijo}} \bibnamefont{and}
  \bibinfo{author}{\bibfnamefont{P.}~\bibnamefont{Singh}},
  \bibinfo{journal}{Phys.Rev.} \textbf{\bibinfo{volume}{D76}},
  \bibinfo{pages}{023510} (\bibinfo{year}{2007}), \eprint{astro-ph/0703566}.

\bibitem[{\citenamefont{Chen et~al.}(2008)\citenamefont{Chen, Wang, and
  Xue}}]{Chen}
\bibinfo{author}{\bibfnamefont{B.}~\bibnamefont{Chen}},
  \bibinfo{author}{\bibfnamefont{Y.}~\bibnamefont{Wang}}, \bibnamefont{and}
  \bibinfo{author}{\bibfnamefont{W.}~\bibnamefont{Xue}},
  \bibinfo{journal}{JCAP} \textbf{\bibinfo{volume}{0805}}, \bibinfo{pages}{014}
  (\bibinfo{year}{2008}), \eprint{0712.2345}.

\bibitem[{\citenamefont{Cai et~al.}(2009)\citenamefont{Cai, Xue, Brandenberger,
  and Zhang}}]{Cai-thermal}
\bibinfo{author}{\bibfnamefont{Y.-F.} \bibnamefont{Cai}},
  \bibinfo{author}{\bibfnamefont{W.}~\bibnamefont{Xue}},
  \bibinfo{author}{\bibfnamefont{R.}~\bibnamefont{Brandenberger}},
  \bibnamefont{and} \bibinfo{author}{\bibfnamefont{X.-m.} \bibnamefont{Zhang}},
  \bibinfo{journal}{JCAP} \textbf{\bibinfo{volume}{0906}}, \bibinfo{pages}{037}
  (\bibinfo{year}{2009}), \eprint{0903.4938}.

\bibitem[{\citenamefont{Biswas et~al.}(2010{\natexlab{a}})\citenamefont{Biswas,
  Mazumdar, and Shafieloo}}]{cyclic-prediction}
\bibinfo{author}{\bibfnamefont{T.}~\bibnamefont{Biswas}},
  \bibinfo{author}{\bibfnamefont{A.}~\bibnamefont{Mazumdar}}, \bibnamefont{and}
  \bibinfo{author}{\bibfnamefont{A.}~\bibnamefont{Shafieloo}},
  \bibinfo{journal}{Phys.Rev.} \textbf{\bibinfo{volume}{D82}},
  \bibinfo{pages}{123517} (\bibinfo{year}{2010}{\natexlab{a}}),
  \eprint{1003.3206}.

\bibitem[{\citenamefont{Biswas et~al.}(2006)\citenamefont{Biswas, Mazumdar, and
  Siegel}}]{BMS}
\bibinfo{author}{\bibfnamefont{T.}~\bibnamefont{Biswas}},
  \bibinfo{author}{\bibfnamefont{A.}~\bibnamefont{Mazumdar}}, \bibnamefont{and}
  \bibinfo{author}{\bibfnamefont{W.}~\bibnamefont{Siegel}},
  \bibinfo{journal}{JCAP} \textbf{\bibinfo{volume}{0603}}, \bibinfo{pages}{009}
  (\bibinfo{year}{2006}), \eprint{hep-th/0508194}.

\bibitem[{\citenamefont{Biswas et~al.}(2007)\citenamefont{Biswas,
  Brandenberger, Mazumdar, and Siegel}}]{BBMS}
\bibinfo{author}{\bibfnamefont{T.}~\bibnamefont{Biswas}},
  \bibinfo{author}{\bibfnamefont{R.}~\bibnamefont{Brandenberger}},
  \bibinfo{author}{\bibfnamefont{A.}~\bibnamefont{Mazumdar}}, \bibnamefont{and}
  \bibinfo{author}{\bibfnamefont{W.}~\bibnamefont{Siegel}},
  \bibinfo{journal}{JCAP} \textbf{\bibinfo{volume}{0712}}, \bibinfo{pages}{011}
  (\bibinfo{year}{2007}), \eprint{hep-th/0610274}.

\bibitem[{\citenamefont{Biswas et~al.}(2010{\natexlab{b}})\citenamefont{Biswas,
  Koivisto, and Mazumdar}}]{BKM}
\bibinfo{author}{\bibfnamefont{T.}~\bibnamefont{Biswas}},
  \bibinfo{author}{\bibfnamefont{T.}~\bibnamefont{Koivisto}}, \bibnamefont{and}
  \bibinfo{author}{\bibfnamefont{A.}~\bibnamefont{Mazumdar}},
  \bibinfo{journal}{JCAP} \textbf{\bibinfo{volume}{1011}}, \bibinfo{pages}{008}
  (\bibinfo{year}{2010}{\natexlab{b}}), \eprint{1005.0590}.

\bibitem[{\citenamefont{Biswas et~al.}(2012{\natexlab{b}})\citenamefont{Biswas,
  Koshelev, Mazumdar, and Vernov}}]{BKMV}
\bibinfo{author}{\bibfnamefont{T.}~\bibnamefont{Biswas}},
  \bibinfo{author}{\bibfnamefont{A.~S.} \bibnamefont{Koshelev}},
  \bibinfo{author}{\bibfnamefont{A.}~\bibnamefont{Mazumdar}}, \bibnamefont{and}
  \bibinfo{author}{\bibfnamefont{S.~Y.} \bibnamefont{Vernov}},
  \bibinfo{journal}{JCAP} \textbf{\bibinfo{volume}{1208}}, \bibinfo{pages}{024}
  (\bibinfo{year}{2012}{\natexlab{b}}), \eprint{1206.6374}.

\bibitem[{\citenamefont{Koshelev}(2013)}]{alex}
\bibinfo{author}{\bibfnamefont{A.~S.} \bibnamefont{Koshelev}}
  (\bibinfo{year}{2013}), \eprint{1302.2140}.

\bibitem[{\citenamefont{Koivisto and Prokopec}(2011)}]{Koivisto:2010pj}
\bibinfo{author}{\bibfnamefont{T.~S.} \bibnamefont{Koivisto}} \bibnamefont{and}
  \bibinfo{author}{\bibfnamefont{T.}~\bibnamefont{Prokopec}},
  \bibinfo{journal}{Phys.Rev.} \textbf{\bibinfo{volume}{D83}},
  \bibinfo{pages}{044015} (\bibinfo{year}{2011}), \eprint{1009.5510}.

\bibitem[{\citenamefont{Ashtekar et~al.}(2006)\citenamefont{Ashtekar,
  Pawlowski, and Singh}}]{Ashtekar}
\bibinfo{author}{\bibfnamefont{A.}~\bibnamefont{Ashtekar}},
  \bibinfo{author}{\bibfnamefont{T.}~\bibnamefont{Pawlowski}},
  \bibnamefont{and} \bibinfo{author}{\bibfnamefont{P.}~\bibnamefont{Singh}},
  \bibinfo{journal}{Phys.Rev.} \textbf{\bibinfo{volume}{D74}},
  \bibinfo{pages}{084003} (\bibinfo{year}{2006}), \eprint{gr-qc/0607039}.

\bibitem[{\citenamefont{Freese et~al.}(2008)\citenamefont{Freese, Brown, and
  Kinney}}]{Freese}
\bibinfo{author}{\bibfnamefont{K.}~\bibnamefont{Freese}},
  \bibinfo{author}{\bibfnamefont{M.~G.} \bibnamefont{Brown}}, \bibnamefont{and}
  \bibinfo{author}{\bibfnamefont{W.~H.} \bibnamefont{Kinney}}
  (\bibinfo{year}{2008}), \eprint{0802.2583}.

\bibitem[{\citenamefont{Baum and Frampton}(2007)}]{Baum}
\bibinfo{author}{\bibfnamefont{L.}~\bibnamefont{Baum}} \bibnamefont{and}
  \bibinfo{author}{\bibfnamefont{P.~H.} \bibnamefont{Frampton}},
  \bibinfo{journal}{Phys.Rev.Lett.} \textbf{\bibinfo{volume}{98}},
  \bibinfo{pages}{071301} (\bibinfo{year}{2007}), \bibinfo{note}{altered and
  improved model astro-ph/0608138 with new title}, \eprint{hep-th/0610213}.

\bibitem[{\citenamefont{Bojowald}(2005)}]{Bojowald:2006da}
\bibinfo{author}{\bibfnamefont{M.}~\bibnamefont{Bojowald}},
  \bibinfo{journal}{Living Rev.Rel.} \textbf{\bibinfo{volume}{8}},
  \bibinfo{pages}{11} (\bibinfo{year}{2005}), \eprint{gr-qc/0601085}.

\bibitem[{\citenamefont{Shtanov and Sahni}(2003)}]{Shtanov}
\bibinfo{author}{\bibfnamefont{Y.}~\bibnamefont{Shtanov}} \bibnamefont{and}
  \bibinfo{author}{\bibfnamefont{V.}~\bibnamefont{Sahni}},
  \bibinfo{journal}{Phys.Lett.} \textbf{\bibinfo{volume}{B557}},
  \bibinfo{pages}{1} (\bibinfo{year}{2003}), \eprint{gr-qc/0208047}.

\bibitem[{\citenamefont{Barragan et~al.}(2009)\citenamefont{Barragan, Olmo, and
  Sanchis-Alepuz}}]{Barragan:2009sq}
\bibinfo{author}{\bibfnamefont{C.}~\bibnamefont{Barragan}},
  \bibinfo{author}{\bibfnamefont{G.~J.} \bibnamefont{Olmo}}, \bibnamefont{and}
  \bibinfo{author}{\bibfnamefont{H.}~\bibnamefont{Sanchis-Alepuz}},
  \bibinfo{journal}{Phys.Rev.} \textbf{\bibinfo{volume}{D80}},
  \bibinfo{pages}{024016} (\bibinfo{year}{2009}), \eprint{0907.0318}.

\bibitem[{\citenamefont{Koivisto}(2010)}]{Koivisto:2010jj}
\bibinfo{author}{\bibfnamefont{T.~S.} \bibnamefont{Koivisto}},
  \bibinfo{journal}{Phys.Rev.} \textbf{\bibinfo{volume}{D82}},
  \bibinfo{pages}{044022} (\bibinfo{year}{2010}), \eprint{1004.4298}.

\bibitem[{\citenamefont{R.C.Tolman}(1934)}]{tolman1}
\bibinfo{author}{\bibnamefont{R.C.Tolman}},
  \emph{\bibinfo{title}{{Relativity,Thermodynamics and Cosmology}}}
  (\bibinfo{publisher}{Oxford U.Press, Clarendon Press}, \bibinfo{year}{1934}).

\bibitem[{\citenamefont{R.C.Tolman}(1931)}]{tolman2}
\bibinfo{author}{\bibnamefont{R.C.Tolman}}, \bibinfo{journal}{Phys.Rev.}
  \textbf{\bibinfo{volume}{37}}, \bibinfo{pages}{1639} (\bibinfo{year}{1931}).

\bibitem[{\citenamefont{Biswas}(2008)}]{emergent-cyclic}
\bibinfo{author}{\bibfnamefont{T.}~\bibnamefont{Biswas}}
  (\bibinfo{year}{2008}), \eprint{0801.1315}.

\bibitem[{\citenamefont{Geshnizjani et~al.}(2011)\citenamefont{Geshnizjani,
  Kinney, and Dizgah}}]{kinney}
\bibinfo{author}{\bibfnamefont{G.}~\bibnamefont{Geshnizjani}},
  \bibinfo{author}{\bibfnamefont{W.~H.} \bibnamefont{Kinney}},
  \bibnamefont{and} \bibinfo{author}{\bibfnamefont{A.~M.}
  \bibnamefont{Dizgah}}, \bibinfo{journal}{JCAP}
  \textbf{\bibinfo{volume}{1111}}, \bibinfo{pages}{049} (\bibinfo{year}{2011}),
  \eprint{1107.1241}.

\bibitem[{\citenamefont{Zinn-Justin}(2001)}]{alpha}
\bibinfo{author}{\bibfnamefont{J.}~\bibnamefont{Zinn-Justin}},
  \bibinfo{journal}{Phys.Rept.} \textbf{\bibinfo{volume}{344}},
  \bibinfo{pages}{159} (\bibinfo{year}{2001}), \eprint{hep-th/0002136}.

\bibitem[{\citenamefont{Ashoorioon et~al.}(2012)\citenamefont{Ashoorioon, Dev,
  and Mazumdar}}]{amjad}
\bibinfo{author}{\bibfnamefont{A.}~\bibnamefont{Ashoorioon}},
  \bibinfo{author}{\bibfnamefont{P.~B.} \bibnamefont{Dev}}, \bibnamefont{and}
  \bibinfo{author}{\bibfnamefont{A.}~\bibnamefont{Mazumdar}}
  (\bibinfo{year}{2012}), \eprint{1211.4678}.

\end{thebibliography}
\end{document}